\documentclass[twocolumn,aps,superscriptaddress,showpacs,nofootinbib,floatfix]{revtex4}
\usepackage{epsfig,bm,feynmf}
\usepackage{graphics}
\usepackage{amsmath}
\usepackage{mathrsfs}
\usepackage[normalem]{ulem}  
\usepackage[dvips]{color} 

\renewcommand\sout{\bgroup \color{red} \ULdepth=-.5ex \ULset}

\newcommand{\bra}[1]{\left\langle #1 \right|}
\newcommand{\ket}[1]{\left| #1 \right\rangle}

\begin{document}

\preprint{APS/123-QED}
\title{Gluon dissociation  of $J/\psi$\ beyond the dipole approximation}

\author{Yunpeng Liu}
\email{yliu@comp.tamu.edu}
 \affiliation{Cyclotron Institute, Texas A$\&$M University, College Station, Texas 77843, USA}

 \author{Che Ming Ko}
 \email{ko@comp.tamu.edu}
\affiliation{Cyclotron Institute, Texas A$\&$M University, College Station, Texas 77843, USA}%
\affiliation{Department of Physics and Astronomy, Texas A$\&$M University, College Station, Texas 77843, USA}%

\author{Taesoo Song}
\email{tsong@comp.tamu.edu}
 \affiliation{Cyclotron Institute, Texas A$\&$M University, College Station, Texas 77843, USA}

\date{\today}

\begin{abstract}
   Using a nonrelativistic potential model, we calculate the cross section for the leading-order gluon dissociation of $J/\psi$ by including the full gluon wave function.  We find that the resulting cross section as a function of  gluon energy is reduced by about a factor of three at its maximum value compared to that calculated in  the dipole approximation that is usually adopted in theoretical studies.   The effect of the reduced cross section on the $J/\psi$  dissociation width is, however, small at finite temperature.
\end{abstract}



\maketitle


\section{Introduction}

The supperssed production of $J/\psi$ in relativistic heavy-ion collisions~\cite{Matsui:1986dk} is one of the most studied probes of the quark gluon plasma (QGP) formed in  these collisions. With experimental data from heavy-ion collisions at various energies~\cite{Gonin:1996wn, Adare:2006ns, Adare:2011yf, Zebo, Abelev:2009qaa, Abelev:2012rv} available from the Super Proton Synchrotron, the Relativistic Heavy Ion Collider, and  the Large Hadron Collider, there have been  extensive theoretical studies to understand  the observed results~\cite{Matsui:1986dk, Xu:1995eb, Capella:1996va, Andronic:2003zv, Thews:2000rj, Lansberg:2006dh, Grandchamp:2003uw, Zhao:2007hh, Yan:2006ve, Song:2011xi}.  Some of these studies were based on the effect of  color screening~\cite{Matsui:1986dk}  or the assumption of statistical production of $J/\psi$ during  hadronization of the QGP~\cite{BraunMunzinger:2000px}.  Other studies also included the effect of $J/\psi$  dissociation by gluons ~\cite{Grandchamp:2001pf, Brambilla:2013dpa} and the inverse process for its production  in the QGP~\cite{Thews:2000rj, Yan:2006ve}. The cross section for the gluon dissociation of $J/\psi$ was first calculated by Bhanot and Peskin using the  operator product expansion~\cite{Peskin:1979va}. This result was later reproduced  using other methods, such as the leading-order perturbative QCD (pQCD) with  the $J/\psi$\ dissociation vertex  obtained from the Bethe-Salpeter equation~\cite{Oh:2001rm}. The latter study was further extended to the next-leading-order pQCD to obtain the higher-order  contribution  to $J/\psi$ dissociation in QGP~\cite{Park:2007zza}. The result of Ref.~\cite{Peskin:1979va} has also been generalized to include the final-state interaction between heavy quarks~\cite{Brezinski:2011ju, Brambilla:2011sg}.  In addition, an effective field theory  based on the potential nonrelativistic QCD  developed in Ref.~\cite{Brambilla:1999xf} provides a systematic way to calculate the width of $J/\psi$ at finite temperature to higher orders.
In all these studies, the dipole approximation that neglects the phase of the gluon wave function or assumes  that the gluon wavelength is much longer than the $J/\psi$ radius. Because the binding energy of $J/\psi$ in vacuum is $\varepsilon=2m_D-m_{J/\psi}=0.64$\ GeV~\cite{Beringer:1900zz} and its radius is $r\approx 0.5$\ fm~\cite{Satz:2005hx}, the phase of the gluon wave function $kr\geq \varepsilon r\approx \pi/2$\ is not  small,  it is thus important to know how the cross section for $J/\psi$ dissociation  by a gluon is affected if the dipole approximation is not used in the calculation.

In the present paper,  the above question is addressed in a nonrelativistic potential model by treating the gluon as an external field. Our results indicate that including the gluon wave function or going beyond the dipole approximation can lead to a factor of three reduction of the $J/\psi$ dissociation cross section and also a reduction of its dissociation width in QGP, which can be appreciable if the potential between the charm and the anticharm quarks is taken to be the internal energy from the lattice QCD calculations.

\section{$J/\psi$ dissociation cross section in a nonrelativistic potential model}

To describe the dissociation of  $J/\psi$ by a gluon, we treat the heavy charm and anticharm quarks in the $J/\psi$ nonrelativistically and the gluon as  a transversely polarized external field using the following Hamiltonian:
\begin{eqnarray}
   H(t) &=& \frac{[{\bf p}_1+g{\bf G}({\bf r}_1,t)]^2}{2m}+\frac{[{\bf p}_2+g{\bf G}({\bf r}_2,t)]^2}{2m}\nonumber\\
   &+&V( {\bf r}_1, {\bf r}_2,t).
\end{eqnarray}
In the above, ${\bf p}_1$ and ${\bf p}_2$ are the canonical momenta of  charm and anticharm quarks; $m$ is their mass; and ${\bf G}({\bf r}_1,t)$ and ${\bf G}({\bf r}_2,t)$ are the gluon fields at the quark and antiquark coordinates ${\bf r}_1$ and ${\bf r}_2$, and $g$ is the QCD coupling constant.  The last term $V({\bf r}_1,{\bf r}_2,t)$ is the potential between  charm and anticharm quarks. The first two terms contain the coupling of the external gluon to the charm quarks, and $V$\ contains that to the exchanged gluon between charm quarks.

In terms of the total momentum ${\bf P}={\bf p}_1+{\bf p}_2$, the relative momentum ${\bf p}=({\bf p}_1-{\bf p}_2)/2$, and the relative coordinate ${\bf r}= {\bf r}_1-{\bf r}_2$ of charm and anticharm quarks, the  Hamiltonian of the interacting charm-anticharm quark pair with an external gluon field to the leading-order in $g$ can be rewritten as
\begin{eqnarray}
   H(t) &=& H_0 + H^\prime(t) \label{hamiltonian},
\end{eqnarray}
with
\begin{eqnarray}
   H_0&=&\frac{ {\bf P}^2}{2M} +\frac{ {\bf p}^2}{2\mu}+ V_0({\bf r}),
\end{eqnarray}
and
\begin{eqnarray}
H^\prime(t)&=&\frac{g}{M} {\bf P}\cdot[{\bf G}({\bf r}_1,t)+{\bf G}({\bf r}_2,t)]\nonumber\\
&+&\frac{g}{m} {\bf p}\cdot[{\bf G}({\bf r}_1,t)-{\bf G}({\bf r}_2,t)]+V_1({\bf r}_1, {\bf r}_2, t).
\end{eqnarray}
 
In the above, $H_0$ is the Hamiltonian of the charm and anticharm quark pair interacting through a Coulomb-like potential $V_0({\bf r})$ with $M=2m$ and $\mu=m/2$  being their total and reduced masses, respectively, and $H^\prime$ is their interaction Hamiltonian with the external gluon, with $V_1({\bf r}_1, {\bf r}_2, t)$ being the leading-order modification to the one-gluon exchange potential $V_0$ owing to the external gluon.

For the case where the gluon has momentum ${\bf k}$ and energy $\omega$, the corresponding gluon field is then
\begin{eqnarray}
   {\bf G}( {\bf x},t)=G^a\mathscr{T}^a{\bm \epsilon}\left(e^{i({\bf k}\cdot {\bf x}-\omega t)}+e^{-i({\bf k}\cdot {\bf x}-\omega t)}\right),
\label{gluon}
\end{eqnarray}
where $G^a$ and $\bm \epsilon$ are its amplitude and polarization vector, respectively,  and $\mathscr{T}^a$  with $a=1, \cdots, 8$ is the $SU(3)$ generator acting on the charm or the anticharm quark that is coupled to the gluon. 

For  a $J/\psi$ at rest,  the contribution from the first term of $H^\prime$ to its dissociation by the gluon vanishes because  its total momentum  is zero.  The matrix element  for a color singlet $J/\psi$ to make a transition to a color octet charm-anticharm quark pair $c{\bar c}$, which are both eigenstates of $H_0$ in Eq.~(\ref{hamiltonian}), after absorbing a gluon then has contributions from the second and third terms of $H^\prime$. For the second term, denoted by $H_I$ in the following, the transition matrix element is given by
\begin{eqnarray}
   &&\bra{c{\bar c}}H_I\ket{J/\psi}\nonumber\\
   &=& \frac{g}{2\mu}\bra{c{\bar c}}{\bf p}\cdot G^a{\bm \epsilon} (e^{i {\bf k}\cdot{\bf r}_1}\mathscr{T}_1^a-e^{i {\bf k}\cdot{\bf r}_2}\mathscr{T}_2^a)\ket{J/\psi},
\end{eqnarray}
where $\mathscr{T}_1^a$\ and $\mathscr{T}_2^a$ are  generators of the fundamental  and conjugate representations of the $SU(3)$\ group in the color space, respectively.  Because the operator $I^a=\mathscr{T}_1^a+\mathscr{T}_2^a$  annihilates a color singlet state $\ket{s}$ and the operator  $J^a=\mathscr{T}_1^a-\mathscr{T}_2^a$  converts a color singlet state to a color octet state $\ket{o}$, i.e. $I^a\ket{s}=0$, and $J^a\ket{s}=C\ket{o}$  with $C=\sqrt{2/N_c}$  if $\ket{s}$ and $\ket{o}$ are normalized to one, we can rewrite the above equation as 
\begin{eqnarray}
   &&\bra{c{\bar c}}H_I\ket{J/\psi}\nonumber\\  
   &=& \frac{gG^a}{4\mu}\bra{c{\bar c}}{\bf p}\cdot {\bm \epsilon} e^{i{\bf k}\cdot{\bf R}}(e^{i {\bf k}\cdot{\bf r}/2}+e^{-i {\bf k}\cdot{\bf r}/2})J^a\ket{J/\psi}\nonumber\\
   &=& \frac{gCG^a}{4\mu}{\bf p}\cdot{\bm \epsilon}\left(\phi( {\bf p}-{\bf k}/2) +\phi( {\bf p}+{\bf k}/2 )\rule{0mm}{4mm}\right)\delta^{(3)}({\bf P}-{\bf k})\nonumber\\
   &=& \frac{g \alpha_{\rm eff}CG^a(2a_0)^{3/2}}{2\pi}\bar{\bf p}\cdot{\bm \epsilon}\frac{\xi^2+(\bar{\bf p}\cdot\bar{\bf k})^2}{\left(\xi^2-(\bar{\bf p}\cdot\bar{\bf k})^2\right)^2}\delta^{(3)}({\bf P}-{\bf k}).\nonumber\\
   \label{eq_shift}
\end{eqnarray}
In the above, ${\bf R}=({\bf r}_1+{\bf r}_2)/2$\ is the center-of-mass coordinate of  charm and anticham quarks, and $\phi({\bf p}\pm{\bf k}/2)$ is the $J/\psi$ wave function in the momentum space. In obtaining the third line  of Eq.~(\ref{eq_shift}), we have neglected the final-state interaction between charm and anticharm quarks, and used plane waves $e^{i{\bf P}\cdot{\bf R}}$ and $e^{i{\bf p}\cdot{\bf r}}$ to describe their center-of-mass and relative motions, respectively.  For the last line of Eq.~(\ref{eq_shift}), it is obtained after we have taken the wave function of  $J/\psi$ to be that of  the one gluon exchange potential $V_0(r)=-\alpha_{\rm eff}/r$, where $\alpha_{\rm eff}=\sqrt{2\varepsilon/\mu}$ is the effective fine structure constant with $\varepsilon$ being the binding energy of $J/\psi$, that is,
\begin{eqnarray}
   \phi({\bf p})=\frac{(2a_0)^{3/2}}{\pi(1+({\bf p}a_0)^2)^{2}},
   \label{eq_wavefunction}
\end{eqnarray}
with $a_0=1/(\mu\alpha_{\rm eff})$  being the Bohr radius.   Also, we have  introduced the dimensionless vectors $\bar{\bf p}={\bf p}a_0$\ and $\bar{\bf k}={\bf k} a_0$\ and the scalar quantity $\xi=1+\bar{p}^2+\bar{k}^2/4$.

\begin{figure}[!hbt]
   \centering
   \includegraphics[width=0.25\textwidth]{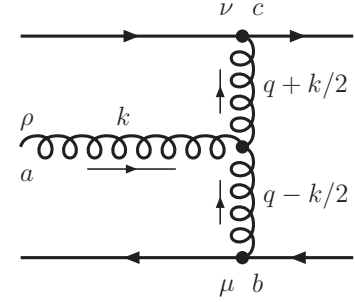}
   \caption{Feynman diagram for heavy quark potential induced by the external gluon field owing to the three-gluon interaction with $a$, $b$, and $c$  denoting the color indices, $\mu$, $\nu$, and $\rho$ denoting the Lorentz indices, and $q$ and $k$ denoting momenta.}
   \label{fig_feynman}
\end{figure}

The last term $V_1$ of $H^\prime$ is the transition potential induced by the absorption of the gluon through the three-gluon interaction shown in Fig.~\ref{fig_feynman}. Using the gluon propagator and the three-gluon vertex given, respectively, by
\begin{eqnarray}
   G^{\mu\nu ;ab}(p)&=&-i\frac{g^{\mu\nu}\delta^{ab}}{p^2+i0^+},\nonumber\\
   \Gamma^{c_1 c_2 c_3}_{\mu_1\mu_2\mu_3}&=& gf^{c_1c_2c_3}[g_{\mu_1\mu_2}(p_1-p_2)_{\mu_3}\nonumber\\
   &+&g_{\mu_2\mu_3}(p_2-p_3)_{\mu_1}+g_{\mu_3\mu_1}(p_3-p_1)_{\mu_2}],
\end{eqnarray}
where $f^{abc}$\ is the structure constant of the $SU(3)$\ group and $p_i$\ is the momentum going into the vertex, the scattering amplitude  for the Feynman diagram in Fig.~\ref{fig_feynman}  is then
\begin{eqnarray}
iM&=&\bar{u}_s(ig\gamma_{\nu})u_s\bar{v}_s(ig\gamma_{\mu})v_su_{cf}^{\dagger}T_1^cu_{ci}v_{ci}^{\dagger}T_2^bv_{cf}\nonumber\\
&\times&G^{\mu\mu';bb'}G^{\nu'\nu;c'c}G^{a}\epsilon^{\rho}\Gamma^{ab'c'}_{\rho \mu'\nu'},
\label{amplitude}   
\end{eqnarray}
where $u_s$\ and $v_s$\ are spinors of charm and anticharm quarks, respectively, and $u_c$\ and  $v_c^*$ are their state vectors  in color space with $T_1^c$  and $T_2^b$ being the conventional half Gell-Mann matrices acting on the color states of charm and anticharm quarks.

The matrix element of the nonrelativistic transition potential $V_1$ is related to the scattering amplitude $M$  by
\begin{eqnarray}
   V_1({\bf q})&\equiv&\bra{ {\bf p}_f, {\bf P}_f \rule{0mm}{3mm}}  V_1 \ket{ \rule{0mm}{3mm} {\bf p}_i, {\bf P}_i } \nonumber\\
   &=& -\frac{1}{(2m)^2(2\pi)^3}M_{fi}\delta^{(3)}({\bf P}_f-{\bf P}_i),
\end{eqnarray}
where ${\bf P}_k$\ and ${\bf p}_k$ are the total and relative momenta of state $k=i,f$, respectively, and ${\bf q}\equiv {\bf 
p}_f-{\bf p}_i$ is the momentum exchanged between the two heavy quarks. Taking the nonrelativistic approximation for the 
spinors $u_s$ and $v_s$, we obtain $\bar{u}_s(\sigma)\gamma_{\nu}u_s(\sigma^\prime)=2m\delta_{\nu 0}\delta_{\sigma
\sigma^\prime}$ and $\bar{v}_s(\sigma)\gamma_{\mu}v_s(\sigma^\prime)=-2m\delta_{\mu 0}\delta_{\sigma\sigma^\prime}$ in the heavy quark limit, where $\sigma$\ and $\sigma^\prime$ are the spin indices, and the $\delta_{\sigma\sigma^\prime}$  implies that the  spins of heavy quarks are not affected by the gluon in the nonrelativistic approximation.  Rewriting $u^{\dagger}_{cf}T_1^cu_{ci}=u^{\dagger}_{cf}\mathscr{T}_1^cu_{ci}
$ and $v^{\dagger}_{ci}T_2^bv_{cf}=v_{cf}^{*\dagger}T_2^{b*}v_{ci}^*=-v_{cf}^{*\dagger}\mathscr{T}_2^{b}v_{ci}^*$  in terms of the previously introduced generators $\mathscr{T}^c$ and $\mathscr{T}^b$  of the fundamental  and  conjugate representations of the $SU(3)$ group  in the color space, respectively, we find \footnote{There are other diagrams contributing to $V_1$ that are related to final-state interactions but are suppressed in the large $N_c$ limit~\cite{Oh:2001rm}. They are not considered here because we are only concerned with the dipole approximation.}
\begin{eqnarray}
V_1({\bf q})&=&ig^2O^{bc}G^{00bb'}G^{00c'c}G^{a}\epsilon^{\rho}\Gamma^{ab'c'}_{\rho 00}
   \delta^{(3)}({\bf P}-{\bf k})/(2\pi)^3,\nonumber\\
\end{eqnarray}
 where 
\begin{eqnarray}
   O^{bc}&\equiv& \mathscr{T}_1^c(-\mathscr{T}_2^b)=-\frac{(I^c+J^c)(I^b-J^b)}{4}.
\end{eqnarray}

Because
\begin{eqnarray}
   O^{bc}f^{abc}&=&-\frac{i}{4}[I^2,J^a],
\end{eqnarray}
where $I^2$\ is the Casimir operator of the $c\bar{c}$ state, the color  operator $O^{bc}$ in $V_1$ transforms a color singlet state to a color octet state. The transition potential $V_1({\bf q})$ can thus be simplified to
\begin{eqnarray}
   V_1({\bf q})&=& \frac{g^3[I^2,J^a]}{2(2\pi)^3}\frac{G^a{\bm \epsilon}\cdot{\bf q}}{ ({\bf q}-{\bf k}/2)^2({\bf q}+{\bf k}/2)^2}\delta^{(3)}({\bf P}-{\bf k}).\nonumber\\
\label{v1}
\end{eqnarray}
The transition  matrix element for $J/\psi$ dissociation by a gluon owing to the three-gluon interaction $V_1$ is then
\begin{eqnarray}
   \bra{c\bar c}V_1\ket{J/\psi}
   &=& \frac{C_A}{2C_F}CG^a\frac{g\alpha_{\rm eff}(2a_0)^{3/2}}{2\pi} {\bm \epsilon}\cdot\bar{\bf p}\nonumber\\
   &\times&\frac{1-(\pi/2-\alpha)\tan\alpha}{\xi^2-\bar{k}^2-(\bar{\bf p}\cdot\bar{\bf k})^2}\delta^{(3)}({\bf P}-{\bf k}),
   \label{eq_V1}
\end{eqnarray}
where $C_A=N_c$ and $C_F=(N_c^2-1)/(2N_c)$ are the Casimir operators in the adjoint and fundamental representations, respectively, $\alpha\equiv\arcsin\left(\bar{k}/\sqrt{\xi^2-(\bar{\bf k}\cdot\bar{\bf p})^2}\right)$, and the relation $\alpha_{\rm eff}=C_F\alpha_s=C_Fg^2/(4\pi)$ has been used.

Adding contributions from $H_I$\ and $V_1$, we find the transition matrix element to be
\begin{eqnarray}
  && \bra{c\bar c}H^\prime\ket{J/\psi}_r = \frac{g \alpha_{\rm eff}CG^a(2a_0)^{3/2}\bar{\bf p}\cdot{\bm \epsilon}}{2\pi}\nonumber\\
  &&\times\bigg(\frac{\xi^2+(\bar{\bf p}\cdot\bar{\bf k})^2}{\left(\xi^2-(\bar{\bf p}\cdot\bar{\bf k})^2\right)^2}
  +\frac{C_A}{2C_F}\frac{1-(\pi/2-\alpha)\tan \alpha}{\xi^2-\bar{k}^2-(\bar{\bf p}\cdot\bar{\bf k})^2}\bigg),\nonumber\\
\end{eqnarray}
where we have dropped the  $\delta$\ function that represents the total momentum conservation and restricted our discussions only to  the relative coordinates, which is indicated by the subscript $r$.\footnote{One can also keep the delta function explicitly, and the result remains the same if the center-of-mass wave function of the $J/\psi$ is properly normalized.}  From the transition rate $w_{J/\psi+g\rightarrow c\bar c}=2\pi|\bra{c\bar c}H^\prime\ket{J/\psi}_r|^2\delta(E_{c\bar c}-E_{J/\psi}-\omega)$ and the gluon current $j=2\omega G^{a2}$,
the cross section for $J/\psi$ dissociation by a gluon is then given by
\begin{eqnarray}
   \sigma&=&  \int \frac{2\pi|\bra{c\bar c}H^\prime\ket{J/\psi}_r|^2\delta(E_{c\bar c}-E_{J/\psi}-\omega)}{2\omega G^{a2}}d{\bf p}\nonumber\\
   &=& \frac{2^7\pi \bar{p}^3}{N_c^2\sqrt{m^3\varepsilon}\xi}\int_{-1}^{1} dx(1-x^2) \bigg(\frac{\xi^2+\bar{p}^2\bar{k}^2x^2}{\left(\xi^2-\bar{p}^2\bar{k}^2x^2\right)^2}\nonumber\\
   &+&\frac{1-(\pi/2-\alpha)\tan \alpha}{\xi^2-\bar{k}^2-\bar{p}^2\bar{k}^2x^2}\bigg)^2,
   \label{eq_beyond_dipole}\\
   \nonumber
\end{eqnarray}
where we have rewritten $\alpha=\arcsin(\bar{k}/\sqrt{\xi^2-\bar{p}^2\bar{k}^2x^2})$ by using the relation $x\equiv\cos\theta=\hat{\bf p}\cdot{\hat{\bm k}}$ and also  the relation $C_A=2C_F$ obtained in the large $N_c$ limit.
The above result reduces to the usual Bhanot and Peskin formula in  the limit $k\rightarrow 0$, corresponding to  the dipole approximation; i.e.,
\begin{eqnarray}
   \sigma^{\textrm{dipole}}&=& \frac{2^7\pi \bar{p}^3}{N_c^2\sqrt{m^3\varepsilon}\xi}\int_{-1}^{1} dx(1-x^2) \left(\frac{1}{\xi^2}+\frac{1}{\xi^2}\right)^2\nonumber\\
   &=& \frac{2^{11}\pi \bar{p}^3}{3N_c^2\sqrt{m^3\varepsilon}\xi^5}.
   \label{eq_peskin}
\end{eqnarray}

We note that the differential cross section for the relative momentum $\bf p$ of final $c\bar c$ to  make an angle $x=\cos\theta$ with respect to the momentum $\bf k$ of the gluon can be simply obtained from Eq.~(\ref{eq_beyond_dipole}) as
\begin{eqnarray}
   \frac{d\sigma}{dx}&=& \frac{2^7\pi \bar{p}^3}{N_c^2\sqrt{m^3\varepsilon}\xi}(1-x^2) \bigg(\frac{\xi^2+\bar{p}^2\bar{k}^2x^2}{\left(\xi^2-\bar{p}^2\bar{k}^2x^2\right)^2}\nonumber\\
   &+&\frac{1-(\pi/2-\alpha)\tan \alpha}{\xi^2-\bar{k}^2-\bar{p}^2\bar{k}^2x^2}\bigg)^2.
\end{eqnarray}

\section{Numerical results and discussions}
\begin{figure}[floatfix]
   \centering
   \includegraphics[width=0.47\textwidth]{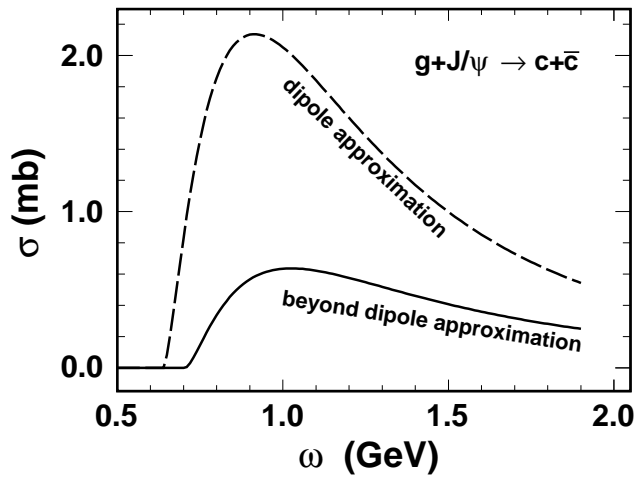}
   \caption{ Cross section for gluon dissociation  of $J/\psi$\ as a function of the gluon energy $\omega$\ in the $J/\psi$\ frame  in (dashed line) and beyond (solid line) the dipole approximation.}
   \label{fig_jpsi}
\end{figure}

We first calculate the total cross section for $J/\psi$ dissociated by a gluon in [see Eq.~(\ref{eq_peskin})] and beyond [see Eq.~(\ref{eq_beyond_dipole})] the dipole approximation  by taking the mass $m_{J/\psi}$\ and the binding energy $\varepsilon$\ of $J/\psi$ as their values in vacuum; that is, $m_{J/\psi}=3.1$\ GeV, $\varepsilon=0.64$\ GeV, and the charm quark mass $m_c=m_D=(m_{J/\psi}+\varepsilon)/2=1.87$\ GeV. The results are shown in Fig.~\ref{fig_jpsi}. It is seen that in the calculation beyond the dipole approximation, the threshold is $0.07$ GeV higher, and the magnitude  of the cross section  is  smaller than that  in the dipole approximation.  The higher threshold energy in the case of beyond the dipole approximation is attributable to the fact that it is $\omega=\varepsilon+{\bf k}^2/2M+{\bf p}^2/2\mu$ and is larger by the amount ${\bf k}^2/2M$ compared to that in the dipole approximation. The smaller cross section from beyond the dipole approximation is partly because it is proportional to $\bar{p}^3$ with $\bar{p}=\sqrt{\xi-1-\bar{k}^2/4}$, as shown in Eq.~(\ref{eq_beyond_dipole}), and $\bar{p}^3$ is smaller than $\bar{p}^3_{\rm dipole}$ with $\bar{p}_{\rm dipole}=\sqrt{\xi-1}$ in the dipole approximation as shown in Eq.~(\ref{eq_peskin}). We note that the dipole approximation is not accurate even for gluon energies near the threshold as a result of the large $J/\psi$\ binding energy.

\begin{figure}[!htb]
   \centering
   \includegraphics[width=0.47\textwidth]{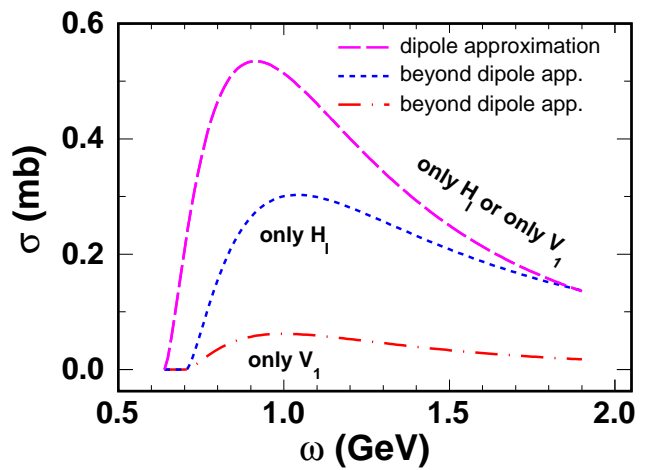}
   \caption{(Color on line)  Cross section for gluon dissociation  of $J/\psi$ as a function of the gluon energy $\omega$ with contributions from only $H_I$ or only $V_1$  in (long dashed line) and beyond [dashed (blue) and dash-dotted  (red) lines] the dipole approximation.}
   \label{fig_quenched_sigma}
\end{figure}

To further understand the above results, we consider separately the contributions from the two terms $H_I$ and $V_1$ in the Hamiltonian.  In the dipole approximation, the two terms give the same cross section  as shown by the long dashed (purple)  line in Fig.~\ref{fig_quenched_sigma}, because they lead to the same transition matrix element as seen in Eq (\ref{eq_peskin}).  Both contributions to the $J/\psi$ dissociation cross section are, however, reduced when they are calculated beyond the dipole approximation as shown by the dashed (blue) and dash-dotted (red) lines in Fig.~\ref{fig_quenched_sigma} for $H_I$ and $V_1$, respectively.

\begin{figure}[!htb]
   \centering
   \includegraphics[width=0.47\textwidth]{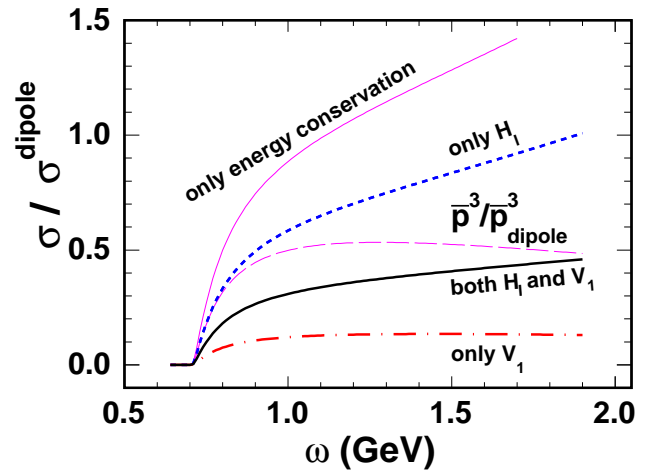}
   \caption{(Color on line)  Ratio of the cross section for gluon dissociation of $J/\psi$ calculated beyond the dipole approximation to that with the dipole approximation. The dashed (blue), dash-dotted  (red) and solid (black)  lines are the ratio with the contribution only from $H_I$, only from $V_1$, and from both, respectively. As a reference, the ratio of $\bar{p}^3$ is also shown as the (purple) long dashed thin  line.}
   \label{fig_sigma_ratio}
\end{figure}

In Fig.~\ref{fig_sigma_ratio}, we show the ratio of the cross section obtained beyond the dipole approximation to that with the dipole approximation.  Also shown by the (purple) long dashed thin  line is the ratio $(\bar{p}/\bar{p}_{\rm dipole})^3$, which is seen to weakly depend on the gluon energy and has a maximum value of 0.53 at gluon energy $\omega=1.28$ GeV. The factor $p^3$ thus contributes at least a factor of 2 to the reduction of the $J/\psi$ dissociation cross section calculated  beyond the dipole approximation.
As implied by Eq.~(\ref{eq_shift}), the  transition matrix element  in the dipole approximation is proportional to the overlap of the initial $J/\psi$ wave function
and the final $c\bar c$ wave function, that is the initial $J/\psi$ wave function $\phi({\bf p})$ in momentum space at final momentum ${\bf p}$. Because of the
inclusion of the gluon momentum ${\bf k}$ in the case of beyond  the dipole approximation, the required relative momentum of $c\bar c$ in $J/\psi$  is ${\bf
p}-{\bf k}/2$\ or ${\bf p}+{\bf k}/2$, corresponding to the absorption of the gluon by the quark  or the antiquark .  
The transition matrix element calculated beyond the dipole approximation is thus proportional to  the $J/\psi$ momentum space wave function  $(\phi({\bf p}-{\bf k}/2)+\phi({\bf p}+{\bf k}/2))/2$, instead of $\phi({\bf p})$ in the dipole approximation.  We note that the value of ${\bf p}$ is reduced when the gluon momentum is taken into account  as a result of
energy conservation. In Fig.~\ref{fig_sigma_ratio}, the (purple)  thin solid line shows the ratio of the cross section owing to only this effect from the energy conservation, that is using the $\phi({\bf p})$ but with $\bf p$ determined with and without including ${\bf k}$. Because the wave function
$\phi({\bf p})$ in Eq.~(\ref{eq_wavefunction}) monotonously decreases with $|{\bf p}|$, the cross section ratio is therefore always above the (purple)
long dashed thin line owing to only the $\bar{p}^3$ factor in the cross section. However, the shift $\pm{\bf k}/2$ of the momentum ${\bf p}$ in the
$J/\psi$ momentum space wave function tends to increase the final-state momentum, thus partly canceling the above effect and resulting in a  final result
for only the term $H_I$ that lies between these two (purple) thin lines as shown by the dashed (blue) thick line. These results
are related to the  fact that the $J/\psi$ wave function $\phi({\bf p})$ decreases monotonously with increasing magnitude of ${\bf p}$
and thus  exists not only for the Coulomb  wave function used in the present study but also for other wave functions, such as a Gaussian wave function, that have a similar behavior.   Therefore, with the inclusion of only $H_I$, the cross section  obtained in the
dipole approximation is larger than that beyond the dipole approximation at small gluon energy $\omega$ but becomes smaller at large gluon energy $
\omega$.

\begin{figure}[!hbt]
   \centering
   \includegraphics[width=0.47\textwidth]{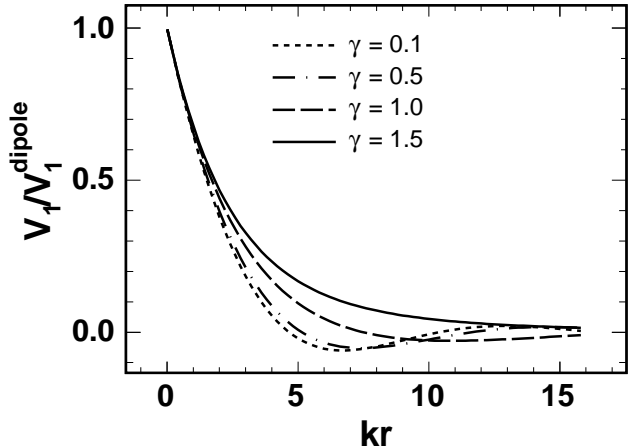}
   \caption{ Ratio of the transition potential $V_1$  owing to the three-gluon interaction to its
   value in the dipole approximation as a function of $kr$  for different values for the angle $\gamma$ between ${\bf k}$ and ${\bf r}$.}
   \label{fig_ratio}
\end{figure}

To understand the reason for the much stronger modification to the transition potential $V_1$ relative to the (purple) thin solid  line owing to only the effect of energy conservation
 on the $J/\psi$ wave function, we rewrite $V_1$ in the coordinate space,
\begin{eqnarray}
   V_1({\bf r})&=& \frac{ig\alpha_s}{4}[I^2,J^a]\frac{{\bf G}^a\cdot{\bf r}}{r}R(kr, \gamma) ,
\end{eqnarray}
where
\begin{eqnarray}
   R(kr,\gamma)&=& \int_0^{\infty}dp\frac{\cos(p\cot\gamma)}{p\kappa \pi\sin\gamma}\bigg(|p-\kappa|K_1(|p-\kappa|)\nonumber\\
   &-&|p+\kappa|K_1(|p+\kappa|)\bigg),
\end{eqnarray}
with $\kappa=kr(\sin\gamma)/2$,  $\gamma$  being the angle between ${\bf k}$ and ${\bf r}$, and  $K$ is the modified Bessel function.  In the dipole approximation, taking $k=0$ gives $R=1$. The potential in  this case is then independent of the magnitude of ${\bf r}$ and thus has a very  long range.  Including the momentum of the external gluon,  the potential is then suppressed for $r\gg 1/k$.  In Fig.~\ref{fig_ratio}, we show the ratio $R=V_1/V_1^{\textrm{dipole}}$ as a function of $kr$  for different values of $\gamma$.  This ratio is seen to decrease dramatically at large $kr$ for all values of $\gamma$,  indicating that the potential $V_1$ does not extend to infinity as in the dipole approximation.  Because of this strong suppression, the cross section owing to the term $H_I$ in the case of beyond the dipole approximation is one order of magnitude smaller than that in the dipole approximation, as shown by the dash-dotted  (red)  thick  line in  Figs.~\ref{fig_quenched_sigma} and~\ref{fig_sigma_ratio}.

Taking both $H_I$\ and $V_1$\ into consideration,  we obtain the ratio of the cross section obtained beyond the dipole approximation to that with the dipole approximation shown as the (black) thick solid  line in Fig.~\ref{fig_sigma_ratio}.  It shows that for gluon energy $\omega\sim 1$ GeV at which the  the $J/\psi$ dissociation
cross section  has the peak value, the ratio is about $0.3$, implying that the dipole approximation overestimates the cross section by a factor of about $3$.

\begin{figure}[!hbt]
   \centering
   \includegraphics[width=0.47\textwidth]{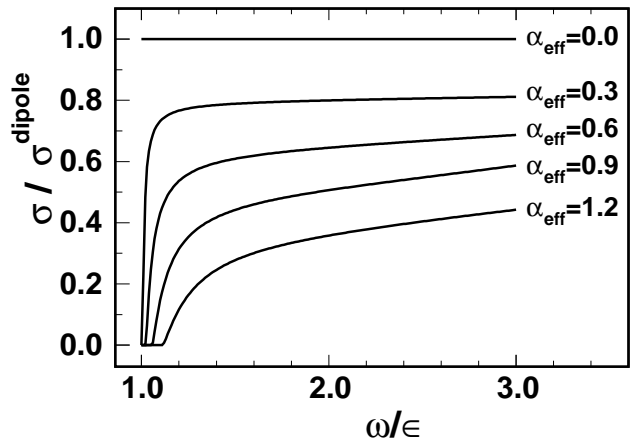}
\caption{ Ratio of  $J/\psi$ dissociation cross section $\sigma$ calculated beyond the dipole approximation to that with the dipole approximation $\sigma^{\textrm{dipole}}$ as a function of the ratio $\omega/\varepsilon$ of the gluon energy $\omega$  in the quarkonium frame and  the $J/\psi$ binding energy $\varepsilon$.}
   \label{fig_dim}
\end{figure}

The cross sections in Eqs.~(\ref{eq_beyond_dipole}) and~(\ref{eq_peskin}) depend on three parameters: the heavy quark mass $m$, the binding energy $\varepsilon$, and the gluon energy $\omega$.  Their ratio  thus depends only on two dimensionless parameters $\omega/\varepsilon$ and $\alpha_{\rm eff}=2\sqrt{\varepsilon/m}$. In Fig.~\ref{fig_dim}, the ratio $\sigma/\sigma^{\textrm{dipole}}$ is shown as a function of $\omega/\varepsilon$ for different values of $\alpha_{\rm eff}$. The ratio is close to one for small values of $\alpha_{\rm eff}$  but is significantly less than one when $\alpha_{\rm eff}\gtrsim 1$ or the energy of the gluon is near the $J/\psi$ dissociation threshold. In our previous calculations, we have used $\varepsilon=0.64$\ GeV and $m=1.87$\ GeV, which gives $\alpha_{\rm eff}\approx1.2$. Similarly, if we consider $\Upsilon$ and take the mass of $b$ quark as that of the lowest $B$ meson, then $m=5.28$ GeV and $\varepsilon=1.1$ GeV, which results in $\alpha_{\rm eff}\approx 0.9$. In both cases, the dipole approximation would overestimate the quarkonium dissociation cross section by more than a factor of two.

\begin{figure}[!hbt]
   \centering
   \includegraphics[width=0.47\textwidth]{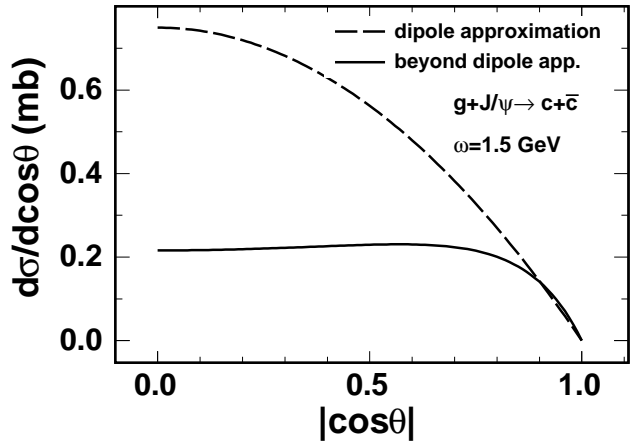}
   \caption{Differential cross section $d\sigma/d\cos\theta$  for the gluon dissociation of $J/\psi$ as a function of $|\cos\theta|$ for the gluon energy $\omega=1.5$\ GeV in the $J/\psi$\ frame, where $\theta$ is the angle between the momentum $\bf k$ of the gluon and the relative momentum ${\bf p}$ of  final $c\bar{c}$.}
   \label{fig_dsigma}
\end{figure}

The momentum of the gluon affects not only the total cross section but also the angular distribution of  final $c\bar{c}$\ pair. Figure~\ref{fig_dsigma} shows the differential cross sections with and beyond the dipole approximation for the gluon energy $\omega=1.5$\ GeV as functions of $|\cos\theta|$ with $\theta$ being the angle between the gluon momentum ${\bf k}$ and the relative momentum ${\bf p}$ of the $c\bar c$ pair.  It is seen that the angular distribution in the dipole approximation decreases with $|\cos\theta|$, and this is because the factor  ${\bm \epsilon}\cdot{\bf p}$ in the transition amplitude [see Eqs.~(\ref{eq_shift}) and~(\ref{eq_V1})], which is the only angular dependence in the dipole approximation, is largest when ${\bf p}$ is parallel to the polarization vector $\bm \epsilon$ that is perpendicular to the momentum ${\bf k}$ of the gluon. Beyond the dipole approximation, the transition amplitude also depends on ${\bf k}\cdot{\bf p}$ [see Eqs.~(\ref{eq_shift}) and~(\ref{eq_V1})], making the relative momentum ${\bf p}$ of final $c\bar{c}$ less likely to be  perpendicular to the momentum of the gluon and thus leading to a smaller differential cross section as shown by the solid line in Fig.~\ref{fig_dsigma}. This is reasonable because after absorbing the gluon the relative momentum of the quarks  is more likely to be in the direction of the gluon momentum. This effect  becomes, however,  smaller when the gluon energy $\omega$ is near the threshold.

\section{$J/\psi$\ width at finite temperature}

\begin{figure}[floatfix]
   \centering
   \includegraphics[width=0.47\textwidth]{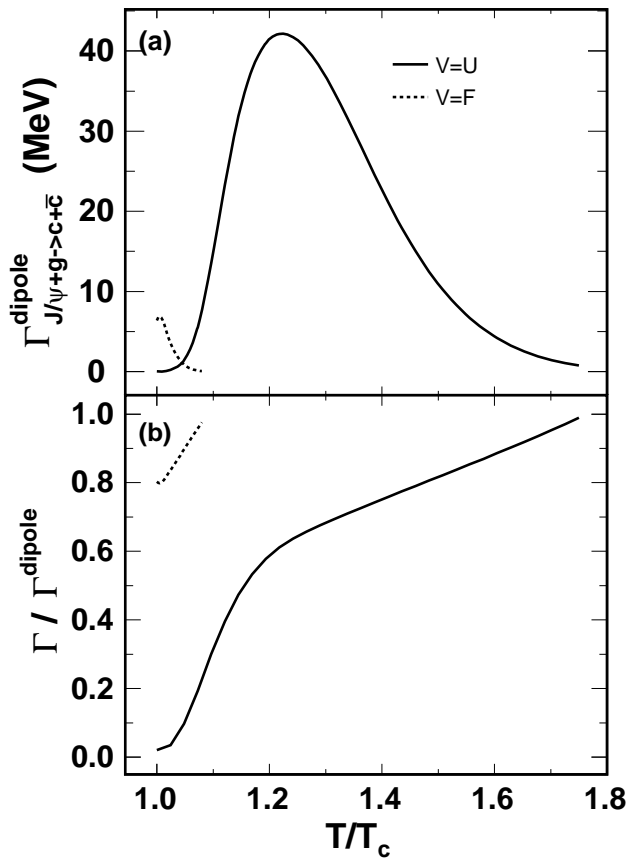}
   \caption{  Temperature dependence of $J/\psi$ width owing to gluon dissociation obtained with the dipole approximation
   (a) and  the inverse of its ratio with respect to that obtained beyond the dipole approximation (b).}
   \label{fig_width}
\end{figure}

To see the effect of $J/\psi$ dissociation cross section obtained beyond the dipole approximation on the $J/\psi$ properties at finite temperature, we compare the  width of $J/\psi$ using this cross section to that in the dipole approximation.  There are two finite temperature effects: finite gluon thermal mass and reduced binding energy of $J/\psi$. For simplicity, we  consider the case that the $J/\psi$ is at rest. The $J/\psi$ width owing to  dissociation by gluons is then
\begin{eqnarray}
   \Gamma(T)=N_g\int\frac{d{\bf k}}{(2\pi)^3}\frac{|{\bf k}|}{\omega (T)}f_g(k, T)\sigma(k,\omega (T)).
\end{eqnarray}
In the above,  $\omega (T)=\sqrt{{\bf k}^2+m_g^2(T)}$ is the gluon energy with the thermal gluon mass $m_g(T)=\sqrt{(2N_c+N_f)/12}gT$, $N_g=24$ is the gluon degeneracy , $f_g$ is the gluon Bose-Einstein distribution, and $\sigma(k, \omega )$ is the gluon dissociation cross section at finite temperature. The latter depends on  the binding energy $\varepsilon$ of $J/\psi$ at finite temperature, which  can be  obtained from solving the Schr\"odinger equation with the heavy quark in-medium potential. For the latter, we use either the free energy $F$ or the internal energy $U=F+TS$ from lattice calculations~\cite{Satz:2005hx, Kaczmarek:1900zz}, for which the $J/\psi$ is bound below $1.15~T_c$ or $2.1~T_c$, respectively, with $T_c=0.165$ GeV being the critical temperature.  Using $N_c=N_f=3$ and $g=2$ in calculating the gluon thermal mass, the $J/\psi$  width calculated in  the dipole approximation  with the $J/\psi$ binding energies obtained from these potentials are shown in the  top panel of Fig.~\ref{fig_width} as a function of the scaled temperature $T/T_c$.  It is seen that the $J/\psi$ width increases with temperature when the temperature is low but decreases with temperature at  high temperature  as a result of decreasing binding energy.

The ratio of the $J/\psi$ width calculated beyond the dipole approximation to that with dipole approximation is shown in the bottom panel of Fig.~\ref{fig_width}. In the free energy  case, the ratio is larger than 0.8 and the dipole approximation is thus reasonable. This is attributable to the small $J/\psi$ binding energy of $110$\ MeV at $T=T_c$ and even smaller values at higher temperature,  which leads to $\alpha_{\rm eff} <0.5$ if the charm mass is taken to be $1.5$ GeV, and the finite gluon thermal mass  makes  the momentum of a gluon  smaller at a given energy.  For the case of using the internal energy as the potential, the width ratio is only somewhat smaller than that of using the free energy as the potential for temperatures higher than $1.2T_c$. This is attributable to the fact that although the binding energy is  large, it is comparable to the mass of the gluon.  For example, the binding energy around the peak at $T=1.2~T_c$ is $0.3\textrm{~GeV}$, and the mass of gluon is about $0.34$ GeV. The $J/\psi$ mass at  this temperature is  $3.3 ~\textrm{GeV}$, which results in $\alpha_{\rm eff} \approx 0.9$.  Because the binding energy decreases monotonously with temperature and $\alpha_{\rm eff}$\ also decreases with temperature,  the ratio becomes  close to $1$  as the temperature approaches that for $J/\psi$ to become unbound.  On the other hand, with decreasing temperature the $J/\psi$ binding energy increases  and the gluon mass becomes smaller, the increased threshold thus requires more energetic gluons.  Because of the smaller number of such gluons, the $J/\psi$ dissociation cross section obtained from beyond the dipole approximation deviates more from that based on the dipole approximation  as the temperature decreases. Because the properties of $J/\psi$, like its mass and binding energy, at finite temperature are still  not very well determined, there is still a large uncertainty in the validity of  the dipole approximation near $T_c$,  although it seems to be a good approximation at high temperature.

\section{Conclusions}

We have derived the cross section for the leading-order gluon dissociation  of $J/\psi$ at zero temperature by including the full gluon wave function. In the limit of long gluon wave length, the usually used cross section  based on the dipole approximation is recovered. In this process, the  gluon can be absorbed by either heavy quarks or by the exchanged gluon between them. Both processes contribute equally in the dipole approximation in the large $N_c$\ limit.  The cross section is, however, smaller beyond the dipole approximation, and this is mainly attributable to the latter process because the momentum of the gluon affects the range of the external gluon induced transition potential.  We have found that the angular distribution of the relative momentum of  heavy quarks in the final state is also modified with the inclusion of the full gluon wave function, leading to heavy quarks more likely to be scattered close to the momentum ${\bf k}$ of the initial gluon rather than perpendicular to it as in the dipole approximation . At finite temperature, the dissociation width of a $J/\psi$ is found to be only slightly modified at high temperature  whether the charm quark potential is taken to be the internal energy or the free energy from the lattice calculations, while the width becomes obviously smaller at low temperature in the case that the internal energy  from the lattice calculations is used as the charm quark potential.

\section*{Acknowledgements}

We thank  Ralf Rapp for helpful discussions. This work was supported by the U.S. National Science Foundation under Grant No. PHY-1068572, the US Department of Energy under Contract No. DE-FG02-10ER41682, and the Welch Foundation under Grant No. A-1358.

\bibliography{ref}

\end{document}